# Transformer-Based Language Models for Similar Text Retrieval and Ranking

**Javed Qadrud-Din   Ashraf Bah Rabiou   Ryan Walker   Ravi Soni   Martin Gajek   Gabriel Pack   Akhil Rangaraj**

Casetext Machine Learning

{javed,ashraf,ryan,ravi,martin,gabe,akhil}@casetext.com

## ABSTRACT

Most approaches for similar text retrieval and ranking with long natural language queries rely at some level on queries and responses having words in common with each other. Recent applications of transformer-based neural language models to text retrieval and ranking problems have been very promising, but still involve a two-step process in which result candidates are first obtained through bag-of-words-based approaches, and then reranked by a neural transformer. In this paper, we introduce novel approaches for effectively applying neural transformer models to similar text retrieval and ranking without an initial bag-of-words-based step. By eliminating the bag-of-words-based step, our approach is able to accurately retrieve and rank results even when they have no non-stopwords in common with the query. We accomplish this by using bidirectional encoder representations from transformers (BERT) to create vectorized representations of sentence-length texts, along with a vector nearest neighbor search index. We demonstrate both supervised and unsupervised means of using BERT to accomplish this task, and show that our approaches outperform several other techniques.

## 1   Introduction

Most existing approaches for retrieving similar text rely in some way on word matching. Word matching based systems fail when texts share the same meaning but have few or zero non-stopwords in common. They also fail when texts that have dissimilar meanings do have many non-stopwords in common (e.g. "Will went to the store" vs "I will store it").

In recent years, machine learning has been used for similar text retrieval by employing models trained to re-rank results originally retrieved by word-match based information retrieval techniques. Such systems often outperform word-match based retrieval alone, but still cannot return results that have no non-stopwords in common with the query text.

Recent advances in neural transformer language model architectures utilizing unsupervised pretraining on large datasets (such as BERT) have made it possible to encode some of the semantic information contained in sentence-length pieces of text. In this paper, we introduce an effective means for using BERT in conjunction with a vector nearest neighbor search index for similar text retrieval and ranking end-to-end, obviating the need for a prior word-match-based information retrieval step.

We present an ablation study in which we demonstrate the relative effectiveness of word-match-based approaches along with various BERT based approaches both with and without different types of task-specific finetuning on a corpus of over 8 million sentences drawn from text in the legal domain (US judicial opinions).

## 2   Background and Related Work

Pretrained neural transformer architectures have been used with success in the past year and a half on a wide range of natural language processing tasks [2], [4], [7], [12], [19], [24]. They are operate by pretraining on large unlabeled datasets and then finetuning to specific labeled tasks. Through the pretraining process, they learn to produce representations of tokens in context that are general enough to form the basis for transfer learning to a wide range of tasks.

Prior to pretrained transformers, Neural Information Retrieval (NIR) models were shown to produce improvements only over weak baselines [22]. However, more recent NIR models based on Contextual Neural Information Retrieval (CNIR) have shown some promise, such as the BERT-based re-rankers proposed by Nogueira & Cho [17], and Yang et. al. [23].

Pretraining approaches prior to pretrained transformers represented each word in one fixed way at inference time, regardless of the specific context in which it is used [9], [15]. Several adaptations have been made to these models for application to information retrieval [5]. But the gains attributed to these non-contextual pretrained models have been shown not to be competitive with strong well-tuned traditional bag-of-words models [22]. We have chosen to focus on CNIR models instead in the hope that they will yield greater gains than non-contextual pretrained models.

Most existing CNIR solutions proceed by first using a bag-of-words model such as BM25 to retrieve the initial set of relevant documents, and then using a BERT-based model to re-rank the top results [3], [13], [17], [23]. This two-step approach has been adopted because the recent transformer-based models are large, and using such a model to directly compare the query with each sentence in a corpus would be too computationally expensive, even if the corpus were relatively small. In our work, we explore the relative effectiveness of several different customized transformer-based models designed to work with vector nearest neighbor search indices. By using vector-based search indices, our techniques need not rely on word matching to generate the initial set of documents, and are thus able to return relevant results that do not necessarily share words with the query. The techniques we explore are as follows:

1. A non-finetuned BERT-based ranker based on TF/IDF weighted averaging of BERT output token vectors coupled with a vector nearest neighbor search index. We call this approach the Unsupervised BERT approach.
2. A BERT-based ranker finetuned to a sentence similarity task using the siamese network sentence similarity approach outlined by Reimers & Gurevynch [20], coupled with a vector nearest neighbor search index.
3. A BERT-based re-ranker, finetuned on query/response pairs in the same manner as the MRPC task in the original BERT paper [4], re-ranking the top results generated by approach number 2 above.
4. Ensemble models that fuse the rankings of one of the BERT-based contextual neural retrieval models with the rankings of bag-of-words models such as BM25 and sequential dependence models.

Additionally, unlike existing CNIR research that focused primarily on short keyword-based queries, our investigation focuses on sentence-length natural language queries. We focus on this type of query because sentence-length queries are useful in similar legal argument retrieval, which is an important task in the legal domain. Sentence-length natural language queries have also been shown to be more challenging for search systems [6]. Hence, tackling this problem in particular is an interesting and challenging endeavor, and likely useful outside the legal domain as well. Using contextual neural models like BERT to tackle this challenge is particularly promising, as recent research has shown that contextual neural models can outperform bag-of-words models on longer natural language queries [3]. BERT is able to capture grammatical structure and word dependencies when processing a natural language query, whereas traditional bag-of-words models do not integrate these key language elements [3].

It has been shown that training sets across different domains can be used to fine-tune contextual neural language models [25]. Similarly, training data designed for one task can sometimes yield improvements when applied to a different task. For example, by leveraging Question Answering data, Yang et al. were able to improve effectiveness on a relevance matching task [23]. In our work, we made use of training data from a different domain to improve our Siamese network-based BERT Ranker model described in Section 3.4.

For vector nearest-neighbor retrieval at scale, various techniques exist for creating tree-like data structures for fast approximate nearest neighbor search. We selected the FAISS library [10] for our investigations.

## 3 Methods

### 3.1 Training and finetuning datasets

The pretraining dataset consists of a large legal domain corpus of all judicial opinions written in the United States above the appeals court level. This dataset consists of 750M sentences.

The model finetuning procedures described below require labeled datasets. In order to generate labeled data for this task, we presented lawyers with query-response sentence pairs. The lawyers rated the pairs using two grades: good or bad match. The pairs were generated by running legal-domain sentence queries through two different sources of results: (1) the Unsupervised BERT Ranker described below and (2) ElasticSearch "more like this", which selects a set of representative terms from the input sentences and executes the resulting query using BM25. These binary good/bad ratings of existing retrieval system responses were used as training data for the BERT Reranker. This was effective for the Reranker because the Reranker need only discriminate between responses that an existing search ranking system generates.

The BERT Ranker model (the Siamese BERT network described in section 3.4 below), however, must discern good results from any possible result in the corpus. To more closely match the task of picking good results from the whole corpus, additional training data was created for the BERT Ranker model in which query sentences were paired with random sentences from the corpus. The training data for the BERT Ranker was thus the same as that for the Reranker except additional random pairs were included. Pairs rated as "good" by humans were labeled "1", pairs humans rated as "bad" were labeled "0.5", and random pairs were labeled "0".

### 3.2 Test data collection

Our test collection consists of 200 queries that we run against a corpus of 8 million sentences drawn at random from text in the legal domain (US judicial opinions). Our task consists in retrieving and ranking sentences that are the most similar to a given query. All of our queries are long natural language queries, each of which is a sentence pertaining to the legal field. The median of the query lengths is 13 terms, whereas the median length of the sentences in the corpus is 26 terms.

To generate query relevance judgments for test data, we ran all our experiments and baselines on the test queries, and pooled the top-10 results from their rankings. We asked lawyers to rate each query-response pair on a 4-grade scale: exactly on-point, relevant, somewhat relevant, and irrelevant. Each query-sentence pair was rated by one assessor. We used these query relevance judgments to compute evaluation measures.

### 3.3 Unsupervised BERT Ranker

BERT employs a masked language modeling approach whereby the model is trained to reconstruct an output sequence from an input sequence in which a set fraction (15%) of tokens is corrupted and/or masked. This approach forces the model to learn language embeddings in an unsupervised fashion from a large unlabeled dataset. The model is further trained using a classification task in which the model predicts whether 2 sentences picked from the training set are consecutive or not. The original BERT as released by google was trained on a combined corpus of about 3.2B tokens based on BooksCorpus and Wikipedia. Mirroring the methodology used to pretrain the original BERT, our model was trained using a sequence length of 128 for 1.3 million steps using a batch of 128 and then the training was finalized using a sequence length of 256 for 44,000 steps.

After pre-training we used a TF/IDF weighted average pooling of the hidden states of the penultimate hidden layer as our sentence embeddings.

We found that the components of BERT token vectors vary widely in scale. This wide scale variance frustrates use of distance metrics for vector comparison, including the cosine distance metric, because certain components dominate the comparison, drowning out others. This type of effect has been noticed in the past in word embedding systems like word2vec [15]. To deal with this effect, we normalized the component scales by dividing each component by its standard deviation.

The i-th sentence output embedding component $y_i$ is given by:

$$y_i = \sum_j TFIDF(j, D) \frac{x_{i,j}}{\sqrt{var_i}}$$

Where j refers to the j-th token in the sentence, $x_{i,j}$ are BERT output embeddings, and $var_i$ is the variance of all i-th embedding components corresponding to all tokens in the dataset.

### 3.4 Fine-tuned BERT Ranker: Siamese-Network Sentence transformer approach

A pretrained model such as BERT can be finetuned for specific tasks in order to improve performance. This typically implies adding further layers on top of the model that convert the output embeddings into useful features for those downstream tasks, and further training the combined model. In the context of text similarity at a scale where pairwise comparison by the model at runtime is computationally infeasible, the model must be trained to produce vector representations of text that can be stored and then compared at runtime using a computationally cheaper distance metric. Finetuning for this task is performed by sequentially supplying the model with pairs of similarity-judgement-labeled sentences and minimizing the cosine loss between the mean pooled output embeddings of said sentences. This approach is depicted in the literature as a "siamese network" where two identical networks with identical and tied weights operate on two distinct input sentences and the loss is computed between the output of those 2 networks [20]. Our model was finetuned on two datasets namely the STSb dataset [1] for 2 epochs and 1 epoch of our custom law dataset described in the third paragraph of Section 3.1.

### 3.5 Reranked Finetuned BERT Ranker

Reranking is a supplemental supervised finetuning procedure in which the model is fitted with a simple binary classification head with binary cross-entropy loss [17]. This classification head is fed via the [CLS] output embedding of the BERT model. The model is then presented simultaneously with pairs of inputs, namely a query sentence and a result sentence under the form <[CLS] query [SEP] result [PAD]> and finetune training is performed by using the binary labels that denote whether or not the result sentence is a good response for the query. The architecture of the full BERT-based similar sentence retrieval and ranking system including the BERT Reranker is illustrated in Figure 1.

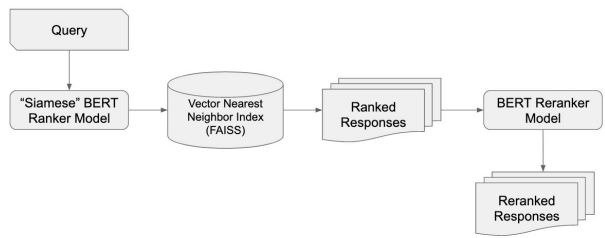

Figure 1: High-level architecture of the full BERT-based similar text retrieval and ranking system. This full system performed best of all our experiments as measured by ncdg@5.

### 3.6 Rank Fusion Method

While supervised methods such as our CNIR methods described above are receiving much attention and producing promising results for ranking and other problems, unsupervised rank fusion models can also be used to improve results. We explored one simple rank fusion model. One benefit of this model is that it is agnostic to the underlying models whose rankings it is fusing. We used it to fuse rankings from our reranked CNIR model described in section 3.5 and our BM25 baseline described in section 4.2. We describe and analyze the results in section 5.

The fusion procedure was as follows. Let D be the set of documents to be ranked, and R be the set of rankings to be fused, each of which is permutation on 1...|D| documents. Then our final fused ranking can be formulated as:

$$Score(d \in D) = \sum_{r \in R} \frac{|D| - pos_r(d) + 1}{|D|} w(r)$$

where $pos_d$ is the position of document d in ranking $r$, and $w(r)$ is the relative weight or importance assigned to ranking $r$. $w(r)$ is a positive integer.

In one of our experiments, we choose $w(r) = 2$ for the reranked CNIR model, and $w(r) = 1$ for the baseline BM25. In another experiment, we set both to 1.

## 4 Experiments

### 4.1 Evaluation Measures

We show nDCG@5 and nDCG@10 results. nDCG, normalized discounted cumulative gain [8] has been used extensively in several information retrieval papers, and is one of the primary measures for several TREC tracks, including Web, Sessions and Microblog tracks. nDCG is a graded relevance measure that rewards documents with high relevance grades and discounts the gains of documents that are ranked at lower positions.

### 4.2 Baselines

We use two baselines. Our first baseline, Okapi BM25, is a TF/IDF-like ranking function based on a probabilistic retrieval framework introduced by Robertson and Jones [21].

Our second baseline is the sequential Dependence Model (SDM) introduced by Metzler and Croft as a Markov Random Field in [14]. SDM is a discriminative probabilistic model that attempts to capture the fact that related terms are likely to appear in close proximity to each other. In our SDM implementation, documents in which adjacent query terms appear in the same order are rewarded, and so are documents in which query terms are in close proximity.

## 5 Results and Discussion

Table 1 shows the results of applying various implementations of context-sensitive neural language models to long natural language queries, against our corpus containing 8 million sentences. The first two rows show the results of our baseline bag-of-words models. We posited that the sequential dependence model, SDM, would improve performance over simple BM25, which does not account for term proximity. However, our results show that BM25 performs better than SDM on long natural language queries. This could suggest that, for queries of this length, accounting for term proximity the way SDM does could be the wrong way of capturing context. For a better understanding of this result, further investigation comparing the effects of SDM on long natural language queries to its effects on short natural language queries and bag of words, would be needed.

|  | ndcg@5 | ndcg@10 |
| --- | --- | --- |
| BM25 | 0.6712 | 0.6609 |
| SDM | 0.6114 | 0.6105 |
| BERT-Reranked BM25 | 0.7375 | 0.7132 |
| Unsupervised BERT | 0.695 | 0.7094 |
| Fine-Tuned BERT | 0.7483 | 0.772 |
| Reranked Fine-Tuned BERT | 0.8082 | **0.8094** |
| Fused Reranked Fine-Tuned BERT x BM25 | **0.8145** | 0.79 |

Table 1: Experimental results applying context-sensitive neural language models to long natural language queries.

The third row in Table 1 shows the results of using BERT to rerank the top-100 results returned by BM25, in a fashion similar to previous studies [23]. As expected, this model leads to significant gains over traditional bag-of-words models.

The fourth row in Table 1 shows that ranking sentences, after first pretraining BERT on the law and then using our Unsupervised BERT approach, leads to an improvement over our strongest bag-of-words baseline (3.5% ndcg@5 increase, and 7% ndcg@10 increase). However, the improvement is small compared to the improvements obtained through all the other BERT-based methods.

The results in rows 5 and 6 of Table 1 present significant gains over both the bag-of-words models and the Unsupervised BERT model. Directly applying Finetuned BERT to rank sentences leads to 11% and 17% improvements in ndcg@5 and ndcg@10 respectively over the strongest baseline. The "Siamese network" Finetuned BERT (which produces results using a vector nearest neighbor search index) also outperforms the model that uses BERT to rerank the top-100 results returned by BM25. Our results suggest that, for long natural language queries, directly ranking sentences using representations created by a finetuned CNIR model yields better results than BERT models that rerank bag-of-words results. This may be explained by the contextual neural language model's ability to retrieve a better initial set of results than a traditional bag-of-words model like BM25.

Furthermore, we find that, starting with an initial list of sentences ranked by the Siamese network fine-tuned BERT-based model, and then reranking that initial list using the BERT

Reranker, leads to further improvements. Reranking led to 8% and 5% increases respectively in ndcg@5 and ndcg@10 over the initial Siamese network-based Finetuned BERT Ranker.

Finally, in the last row of Table 1, we show the results of the Rank Fusion (RF) model that gives more importance to the Reranked Fine-Tuned BERT ranker ($w(r) = 2$ for the Reranked Fine-Tuned BERT ranker model, and $w(r) = 1$ for the baseline BM25 model). The results are close, but it is worth noting that ndcg@5 results for the RF model are higher than both models that are being fused, suggesting that the RF model may be a good way to promote better sentences to the top of the ranked list.

## 4 Conclusion and Future Work

In this paper, we showed that we can obtain substantial gains in ranking effectiveness for long natural language queries by making modifications to a contextual neural language model, BERT. Additionally, we showed that directly using a finetuned BERT model with a "siamese network" architecture to rank sentences outperforms using BERT to rerank an initial list of sentences retrieved by BM25. We obtained even further gains when using a finetuned BERT-based reranker to rerank sentences that were initially ranked by the BERT-based ranker.

Interesting future work includes investigating and comparing these BERT-based ranking models not just on long natural language queries, but also on queries of varied lengths. Additionally, the fact that these BERT-based methods work well suggests that more recent neural transformer based language models, which have been shown to outperform BERT across a range of natural language tasks [2], [7], [12], [19], [24], may yield further gains for similar sentence retrieval and ranking.